\documentclass[nofootinbib,superscriptaddress,onecolumn,showpacs]{revtex4}
\usepackage{graphicx}
\usepackage[english]{babel}
\usepackage{amsmath,amssymb}

\newcommand{\ie}{\emph{i.e., }}
\newcommand{\reff}[1]{(\ref{#1})}
\newcommand{\eref}[1]{Eq.\reff{#1}}
\newcommand{\erefs}[1]{Eqs.\reff{#1}}
\newcommand{\figref}[1]{FIG.\ref{#1}}
\newcommand{\citer}[1]{Ref.\cite{#1}}
\newcommand{\citers}[1]{Refs.\cite{#1}}
\newcommand{\omp}{\omega_p}
\newcommand{\p}{\partial}
\usepackage{color}

\begin{document}
\title{Quasi-linear model for the beam-plasma instability:\\
       analysis of the self-consistent evolution}

\author{Giovanni Montani}
\affiliation{ ENEA, Fusion and Nuclear Safety Department, C. R. Frascati,\\
              Via E. Fermi 45, 00044 Frascati (Roma), Italy}       
\affiliation{ Physics Department, ``Sapienza'' University of Rome,\\
              P.le Aldo Moro 5, 00185 Roma, Italy}

\author{Francesco Cianfrani}
\affiliation{University of Rome Tor Vergata, Department of Industrial Engineering, \\Via del Politecnico 1, Rome 00133, Italy}

\author{Nakia Carlevaro}
\affiliation{ ENEA, Fusion and Nuclear Safety Department, C. R. Frascati,\\
              Via E. Fermi 45, 00044 Frascati (Roma), Italy}
\affiliation{Consorzio RFX, 
Corso Stati Uniti 4, 35127 Padova, Italy}

%\date{}

\begin{abstract}
We re-analyze the quasi-linear self consistent dynamics for the beam-plasma instability, by comparing the theory predictions to numerical simulations of the corresponding Hamiltonian system. While the diffusive features of the asymptotic dynamics are reliably predicted, the early temporal mesoscale transport appears less efficient in reproducing the convective feature of the self-consistent scenario. As a result, we identify the origin of the observed discrepancy in the underlying quasi-linear model assumption that the distribution function is quasi-stationary. Furthermore, we provide a correction to the instantaneous quasi-linear growth rate based on a linear expansion of the distribution function time dependence, and we successfully test this revised formulation for the spectral evolution during the temporal mesoscale.
\end{abstract}
\pacs{52.25.Dg; 52.40.Mj; 52.35.Mw}
\maketitle

\section{Introduction} 
One of the most useful paradigm of the theoretical analysis of Plasma Physics is the so-call bump-on-tail (BoT) scheme, describing the instability induced in a thermal plasma due to a suprathermal population of charged particles \cite{ZCrmp}. The relevance the BoT paradigm relies on its capability to capture the main features of the fast particle radial transport in Tokamak devices, due to a one-to-one correspondence between the gradients in the velocity distribution finction and the pressure gradients in the corresponding radial direction \cite{BB90a,BB90b,BB90c}. As shown in \citers{EEbook,AEE98,EE08} and in \citers{CFMZJPP,ncentropy}, the BoT scenario can be suitably described via a beam-plasma system (BPS) model by a cold background dielectric on which fast electrons interact with the longitudinal Langmuir waves, as first postulated in \citers{OM68,OWM71}, see also \citers{MK78,TMM94,KV14} and \citer{L72} for the case of a warm beam.

We will use a $N$-body Hamiltonian BPS formulation for testing the predictivity of the quasi-liner (QL) dynamics. When the Langmuir spectrum with which the tenuous fast electron beam interacts is sufficiently continuous and large (in comparison to the non-linear velocity spread), it is well-known, since from the original treatment \cite{Pines} (see also the detailed monographs \citers{MT64,KT73,D82}), that the transport in the velocity space acquires a marked diffusive character and the particle distribution function obey a Fokker-Planck equation, whose diffusion coefficient is proportional to the electric field spectrum intensity. This behavior accounts for a random walk of the electron velocity, due to the continuous scatter with the superposition of many different wave components, having a random phase distribution \cite{pesme94}. Since the beam is tenuous, the electric field amplitude is to be regarded as small even after its saturation and it constitutes a small control parameter to ensure the consistency of the QL approximation, especially about the possibility to neglect cross coupling terms in the wave amplitudes. For a satisfactory discussion about the validity of the QL model and its limit of application, see \citer{Laval99} and refs. therein. While an implementation of the QL diffusive scenario to describe radial transport in Tokamak devices is given in \citer{GG12}, where the so-called $1.5$ model is addressed. Morever, in \citer{GBG14} it is provided a re-analysis of the Line Broadened QL model \cite{BBFW95} and a comparison with respect to a fully Vlasov solver for the one dimensional reduced scheme in the presence of collisional effects and focusing on the case of isolated modes.

Clearly, the pure diffusive QL model properly accounts for the BoT evolution only in sufficiently late phases when each particle is, on average, scattering with all the field components simultaneously and no significant trapping processes take place. As it has been outlined in \citer{ncentropy}, such a situation is not appropriate to describe the temporal mesoscale evolution, where convective transport can be important. In fact, in this intermediate region of the evolution, a significant fraction of particles can coherently interact with a finite number of wave components, giving rise to ballistic motion in the velocity space and avalanche phenomena. For the relevance of such features in the radial transport of fast ions in ITER relevant simulations, see \citer{spb16}. In \citer{ncentropy}, it has been emphasized that trace of such coherent evolution of the particle-field interaction can be found even in the QL dynamics, as far as the self-consistent evolution is accounted to some extent.

In this paper, we focus our attention just to the emergence of such non-diffusive transport in the early non-linear phases of the BPS, in order to evaluate the predictivity of the QL paradigm in the temporal mesoscale, when structures in the phase space (like clumps and holes) can still survive. To this end, after a qualitative introduction of the problem, we write down the QL model equations, in particular the single partial differential equation (PDE) for the spectrum evolution.
% We numerically solve such an equation and we calculate the electron distribution function at different instants, from the field linear growth and saturation, up to the later diffusive stages. We recover the formation of a flattening region in the velocity profile, associated to a zero drive of the inverse Landau damping.}
and, in order to characterize the prediction of the QL theory, we compare the profiles of the calculated distribution function with those ones obtained by the Hamiltonian $N$-body BPS \cite{CFMZJPP,ncentropy}. In this work, we adopt such a description which directly follows from the original paper \citer{OWM71}. It is worth mention that also particle-in-cell (PIC) numerical simulations of the BoT instability is widely addressed in plasma physics (see, for example, the seminal papers \citers{DS68,MN69,D90}). As well known, PIC codes are based on a Lagrangian frame and sample the plasma density as a series of macro-particles, which represent parcels of density. Particles themselves are assigned a weighting, which represents the amount of phase-space density represented by the particle, and a position in phase space (details of this approach can be found in \citers{H55,HE88}). For more recent applications, see, \emph{e.g.}, \citers{D94,EB15,BDE17} and refs therein, and \citer{DF14}. The most significant discrepancy, emerging in the above discussed comparison, concerns the system time ``reactivity'', \ie the QL dynamics appears as slower with respect the exact Hamiltonian one (see also \citer{LP83}). While the non-linear velocity spread is qualitatively reproduced, say with enough precision in view of the radial Tokamak transport (see the analysis in \citer{nceps16}), the achievement of the plateau configuration and, on average, the whole mesoscale evolution appear a bit delayed in the QL scenario.

The main merit of the present analysis consists of giving a clear explanation for such a slower response of the QL model in comparison to a real BPS, in terms of the quasi-stationary assumption at the ground of the derivation of a pure diffusive dynamics. In the early non-liner stages of the BPS, bouncing-like processes, as discussed in \citer{OWM71}, take place and the dynamics of the distribution function is associated to the same time scale of the electric field: in such a regime, the quasi-stationary assumption must be relaxed, or at least improved, to get a more realistic picture of the system, as it appears in the Hamiltonian model. This information is of a significant impact when, like in \citer{nceps16}, we want to predict the radial transport of fast ions in Tokamak configuration via a one-to-one mapping from the velocity to the physical space. While, as shown in \citer{GG12} the prediction of the diffusive QL evolution are of interest for the radial transport, here we suggest that the mesoscale evolution can not be reliably mapped before a suitable upgrade of the QL self-consistent picture is reached.

We make a first step to upgrade the QL model by considering a more refined approximation scheme for the quasi-stationary distribution function, in which a linear in time contribution is retained. As shown by using the numerical distribution function, the modified growth rate is enhanced with respect to the QL case, implying that the system reactivity increases. Hence, a better agreement with simulations, thus a better characterization of the mesoscale dynamics, is achieved.

The paper is organized as follows. In Sec.\ref{sec2}, we provide some basic concepts introduced via a qualitative picture of the problem. Furthermore, using the fundamental hypotheses of the QL model, we arrive to the basic equation for the electric field spectral evolution, which constitutes the tool to test the self-consistency of the QL paradigm. In Sec.\ref{sec6}, we introduce a schematic description of the Hamiltonian BPS and a discussion of the comparison with the QL prediction for the distribution function morphology is provided. In Sec.\ref{sec7}, the growth rate is explicitly evaluated by relaxing the quasi-stationary assumption for the distribution function, and tests for the reliability of the predicted dynamics are numerically provided. In Sec.\ref{sec8}, concluding remarks follows. %In Sec.\ref{appendix}, we derive the Poisson equation for the BPS by stressing how the concept of instantaneous frequency concerns both the field evolution and the dielectric morphology.

\section{Quasi-linear dynamics}\label{sec2}
We focus on a warm beam of electrons, having an initial homogeneous spatial distribution with number density $n_B$, interacting with a cold homogeneous plasma of density $n_p\gg n_B$. The dynamics of a beam in the sea of Langmuir waves present in the plasma is faced on a 1D level. The addressed electric field $E(x,t)$ is expressed using the Fourier decomposition $E(t,x)=\sum E_k(t) e^{ikx}$, where $k$ denote the considered wave numbers for the Langmuir spectrum.

Those electrons which have a velocity $v$ very close to the phase velocity of an electric wave component $v_{Ph}=\omega_p/k$ ($\omega_p$ being the plasma frequency, $\omp^2=4\pi n_p e^2/m$, where $e$ and $m$ denote the electron charge and mass, respectively) can efficiently exchange power with such a mode, resulting trapped in the local electric well. In this sense, let us introduce the trapping (bounce) frequency $\omega_B$ defined by $
\omega_B^2 \equiv 2 ek|E_k^S|/m$, where $|E_k^S|$ denotes the saturated values of $E_k$. In fact, after a linear growth due to the inverse Landau damping process \cite{oneil65}, the electric field reaches a saturated value corresponding to the maximum energy extraction from the resonant electrons.

The time scale $\tau_B=2\pi/\omega_B$ characterizes the typical trapping time of an electron within the potential. We did not label $\omega_B$ by the suffix $k$, because, in what follows, we will intend to deal with an average value, estimated with an average electric field saturation level or with the most relevant Fourier component. Another important time scale is the autocorrelation time $\tau_{ac}$, \ie the typical time an electron takes to cross the spatial size $\Delta x$ of the wave packet, namely $\tau_{ac}=\Delta x/v = 1/(v\Delta k)$, $\Delta k$ being the packet spread in wavenumbers. Also in this estimate, the velocity must be intended as the typical resonant one of the beam. Clearly, the trapping of a large number of electrons in the electric field, associated to the formation of structures in the phase space, can take place only if $\tau_B\ll\tau_{ac}$. In such a situation, resonant particles can get trapped by a small set of modes before fell the whole fluctuation spectrum. Otherwise, for $\tau_{B}\gg\tau_{ac}$, the particles feel simultaneously all the field components and a diffusion-like process takes place.

The random character of the Langmuir electrostatic field leads us to require that its mean value, taken over a sufficiently large time, larger than $\tau_{ac}$, essentially vanishes, \ie $\langle E(t,x)\rangle \simeq 0$. Furthermore, because of the average spatial homogeneity of the considered system, the variance of the electric field $\langle E^2\rangle$ must depend on time only: the weak turbulence arising in the system does not produce structures on spatial scales comparable to the fundamental one $L$ (here $L$ deotes the system periodicity spatial size, \ie $k= 2\pi \ell /L$, $\ell$ being an integer number). If we deal with a Langmuir continuum ($L$ is sufficiently large), we can evaluate the variance obtaining the basic condition
\begin{equation} 
\langle E_k E_{k^{\prime}}\rangle = |E_k|^2 \delta (k+k^{\prime})\;,
\label{ql6} 
\end{equation}
where $\delta$ denotes the Dirac function. This issue clearly suggests that, as far as macroscopic homogeneity is preserved, no significant coupling can take place, on average, between Fourier components associated to different wavenumbers. Such a consideration is at the ground of the QL model.

Furthermore, in the limit $\tau_B\gg\tau_{ac}$, the superposition of different harmonics interacting with the same electron allows us to model the electric field as a white noise. Thus, the force equation of the system can be written in the form of a random walk in the velocity space, namely described by the Langevin equation $dv/dt=-e E(t,v)/m$, where the dependence on the particle velocity is inferred by means of the resonance condition, which selects certain groups of harmonics, while the electric field satisfies the condition
\begin{equation} 
\langle E(t,v) \rangle = 0 \;, \qquad 
\langle E(t,v)E(t+\tau,v) \rangle = \Delta(v)\delta(\tau)\;.
\label{ql8} 
\end{equation} 
where $\Delta(v)$ is a given function of the velocity. By other words, when the electron interacts with a large number of Fourier harmonics having random distributed phases, we deal with a white noise in the electron velocity space. It is well-known that the probability $d\mu$ of finding an electron in the range ($v$, $v+dv$) is $d\mu=P(t,v)dv$, where the probability density $P$ satisfies a Fokker-Planck equation of the form
\begin{equation} 
\p_t P =\p_v (\mathcal{D}(v)\p_v P)\;,
\label{ql9} 
\end{equation}
where $\mathcal{D}=e^2 \Delta/2m^2$. This picture closely reproduces some hypotheses and the main predictions of the QL model in the limit of sufficiently large times with respect to $\tau_{ac}$. However, we stress how the condition \reff{ql6} (spatial homogeneity of the field) is much general than the more restrictive assumption \reff{ql8} (Gaussian field) responsible for a diffusive behavior. In this paper, we are interested to study the self-consistent dynamics allowed by the preservation of the system homogeneity, in order to characterize the evolution from the initial linear instability phase, up to the diffusive regime.

\subsection{QL equations}\label{sec4}
The QL model is derived from the Vlasov-Poisson coupled system specified for the the interaction of the suprathermal electron beam with a cold background plasma and it is expressed, in Fourier space, in terms of $E_{k}(t)$ and the distribution function components $f_k(t,v)$ of the beam particles. The attention is focused on behavior of $f_{k=0}\equiv f_0$, which is the only contribution having a non-zero initial condition since the BPS is initially spatially homogeneous. The dynamics of $f_0$ receives contributions from all $k$-values and, thus, it accounts for the main transport properties of the system. Moreover, according to the homogeneity condition \eref{ql6}, the component $f_k$ is assumed to receive mainly contribution only by the correspondent harmonics. Introducing the assumption that the function $f_0$ is not fast varying in time (quasi-stationarity hypothesis associated to the plateau formation), the QL model equations can be written as follows \cite{Pines,LP81}:
\begin{align}
&\p_t f_0(t,v) = \p_v(\mathcal{D}_{QL}\p_v f_0)\;,\label{st1}\\
&\p_t |E_k(t)|^2 = 2\gamma^{QL}_k |E_k|^2\;,\label{sv}
\end{align}
where $\mathcal{D}_{QL}(t,v)$ and $\gamma^{QL}_k(t)$ are the QL diffusion coefficient and instantaneous growth rate, respectively, and read
\begin{align}
&\mathcal{D}_{QL}(t,v) \equiv \frac{e^2\pi\mathcal{N}}{m^2v}|E_k|^2 \big|_{k=\omp/v}\;,\label{st2}\\
&\gamma^{QL}_k(t) \equiv\frac{2\pi^2e^2\omega _p}{mk^2}\p_v f_0 \big|_{v=\omp/k}\;,\label{svGamma}
\end{align} 
here we have used the resonance condition $kv=\omp$ and since the model requires a sufficiently large spectral density, we have approximated the discrete $k$-space by a continuum. In fact, we have introduced $\mathcal{N}(k)$ as the state density and it can be assumed constant $\mathcal{N}\simeq K/\Delta k$, where $K$ is the number of the considered modes while $\Delta k=k_{max} - k_{min}$ indicates the spectral width (the interval where the spectral intensity is significantly different from zero). Out of $\Delta k$, both the spectral density and intensity vanish.

The electric field and the distribution function dynamics can be combined (see also \citer{ncentropy}) to get a partial differential equation in the spectral intensity only. First of all, by means of the resonance condition and the consideration above, we move to the velocity space introducing the electric potential modes as $E_k(t)\to-i\omp\varphi(t,v)/v$, where $\varphi(t,v)$ has to be intended as the continuous mode spectrum derived from the discrete one, specified for $v=\omp/k$. In order to compare the QL model with respect to numerical simulations, let us now introduce the following normalization. The 1D cold plasma equilibrium is taken as a periodic slab of length $L$ while time is taken in $\omp$ units $\tau=t\omp$. We also introduce the beam to plasma density ratio as $\eta=n_B/n_p$. Other quantities are defined as as: $\ell=k(2\pi/L)^{-1}$ (integer numbers), $u=v(2\pi/L)/\omp$, $\phi=(2\pi/L)^2 e\varphi/m\omp^2$. In this scheme, we analyze the spectral intensity $\mathcal{J}(\tau,u)=|\phi|^2$ and we use the beam distribution function $f_B$ defined by transforming $f_0(t,v)/n_B\to f_B(\tau,u)$. The QL equations write now as
\begin{align}\label{QL_u}
f_B=F_{B}+\bar{\mathcal{N}} M \eta^{-1}\;\p_u(\mathcal{J} u^{-5})\;,\\
\p_{\tau}\mathcal{J}=\pi\eta M^{-1} u^2 \mathcal{J} \p_u F_{B}+ \pi \bar{\mathcal{N}} u^2 \mathcal{J} \p_u^2 (\mathcal{J} u^{-5})\;,
\label{spect}
\end{align}
where $F_{B}(u)$ indicates the initial beam profile in the $u$ space, $\bar{\mathcal{N}}=m/\Delta\ell$ (here $\Delta\ell=\ell_{max}-\ell_{min}$) and $M=\int_{-\infty}^{+\infty} du F_{B}$.

\section{Comparison with numerical simulations}\label{sec6}

We now compare the QL distribution function and spectrum evolutions with those ones emerging from numerical simulations of an $N$-body Hamiltonian system describing the BPS \cite{CFMZJPP,ncentropy}.

From the linear theory, the dispersion relation for a single mode having corresponding (normalized) resonant velocity $u_\ell=1/\ell$ reads as \cite{OM68,LP81}
\begin{align}\label{disrel}
2(\bar{\omega}_{0\ell}+i\bar{\gamma}_{\ell}-1)-\frac{\eta u_\ell}{M}
\int_{-\infty}^{+\infty}\!\!\!\!\!\!\!du\frac{\p_u F_B}{u/u_\ell-\bar{\omega}_{0\ell}-i\bar{\gamma}_\ell}=0\;,
\end{align}
(for an analysis on the accuracy of this equation, see \citer{nceps18}). Here, the dielectric function is expanded near $\omega_\ell\simeq\omp$ (according to motion equations, see below) and we used $\bar{\omega}_\ell=\bar{\omega}_{0\ell}+i\bar{\gamma}_\ell$, where $\bar{\omega}_{0\ell}$ denotes the real part of the frequency and $\bar{\gamma}_\ell$ ($\ll1$) the growth rate of the considered mode (the bar indicates that these quantities are in $\omp$ units). The linearized form of \eref{disrel} (by assuming $\bar{\omega}_{0\ell}=1$) reduces to the well know expression
\begin{align}\label{drlin}
\bar{\gamma}_{L\ell}=\frac{1}{2}\,\pi\eta M^{-1}u_\ell^2\p_u F_B\big|_{u_\ell}\;,
\end{align}
which of course corresponds to the steady form \eref{svGamma} and, by construction, it is the discretization of the spectral growth rate of \eref{spect} (first term of the right-hand side). It is important to notice that a break-down of the perturbative Landau damping expression (\eref{drlin}) is observed (compared with respect to numerical integration of \eref{disrel}) for specific setup of the beam plasma system, enlightening non-perturbative and non-local character of the dispersion relation, especially in the flat region of the initial distribution function.

In what follows, we careful choose the initial conditions of the system so that the numerically evolved modes have growth rate values well predicted by the Landau expression. This way, we ensure that discrepancies in the system evolution between the numerical simulation and QL model do not rely on different drive values. In fact, the QL theory naturally defines an instantaneous Landau growth rate as far as the field amplitude is small enough.

\subsection{N-body beam plasma system description}
We adopt the Hamiltonian formulation of the BPS \cite{CFMZJPP,ncentropy}, where the broad fast particle beam self-consistently evolves in the presence of several modes at the plasma frequency. This ensures a nearly vanishing dielectric function of the cold background plasma ($\epsilon=1-\omp^2/\omega^2$) and allows casting the Poisson equation for each mode into the form of an evolution equation (while particle trajectories are solved from the force equation). Each beam particle position along the $x$ direction is labeled by $x_i$ ($N$ indicates the total particle number) and  we use the normalization introduced in the previous Section (positions are scaled as $\bar{x}_i=x_i(2\pi/L)$). The BPS is governed by the following set of equations:
\begin{equation}\label{mainsys1}
\begin{split}
&\p_{\tau}\bar{x}_i=u_i \;,\\
& \p_{\tau} u_i=\sum_{\ell}\big(i\,\ell\;\bar{\phi}_\ell\;e^{i\ell\bar{x}_{i}}+c.c.\big)\;,\\[-15pt]
&\p_{\tau} \phi_\ell=-i\phi_\ell+\frac{i\eta}{2\ell^2 N}\sum_{i=1}^{N} e^{-i\ell\bar{x}_{i}}\;,
\end{split}
\end{equation}
and we recall that resonance conditions rewrite as $\ell u_\ell=1$. In the non-linear simulations, we initialize the beam particle numbers distributed according to a given $F_B(u)$ (for a total $N=10^{6}$ particles). We solve \erefs{mainsys1} using a Runge-Kutta (4th order) algorithm and the initial conditions for the particle positions $\bar{x}_i$ are given uniformly in $[0,\,2\pi]$, while modes are initialized at amplitudes of order $\mathcal{O}(10^{-14})$ to ensure the starting linear regime. The resulting evolutive velocity distribution function is simply evaluated by a smoothed $x$-dimension averaged procedure of the correspondent phase-space.

\begin{figure}[ht!]
\centering
\includegraphics[width=.4\textwidth,clip]{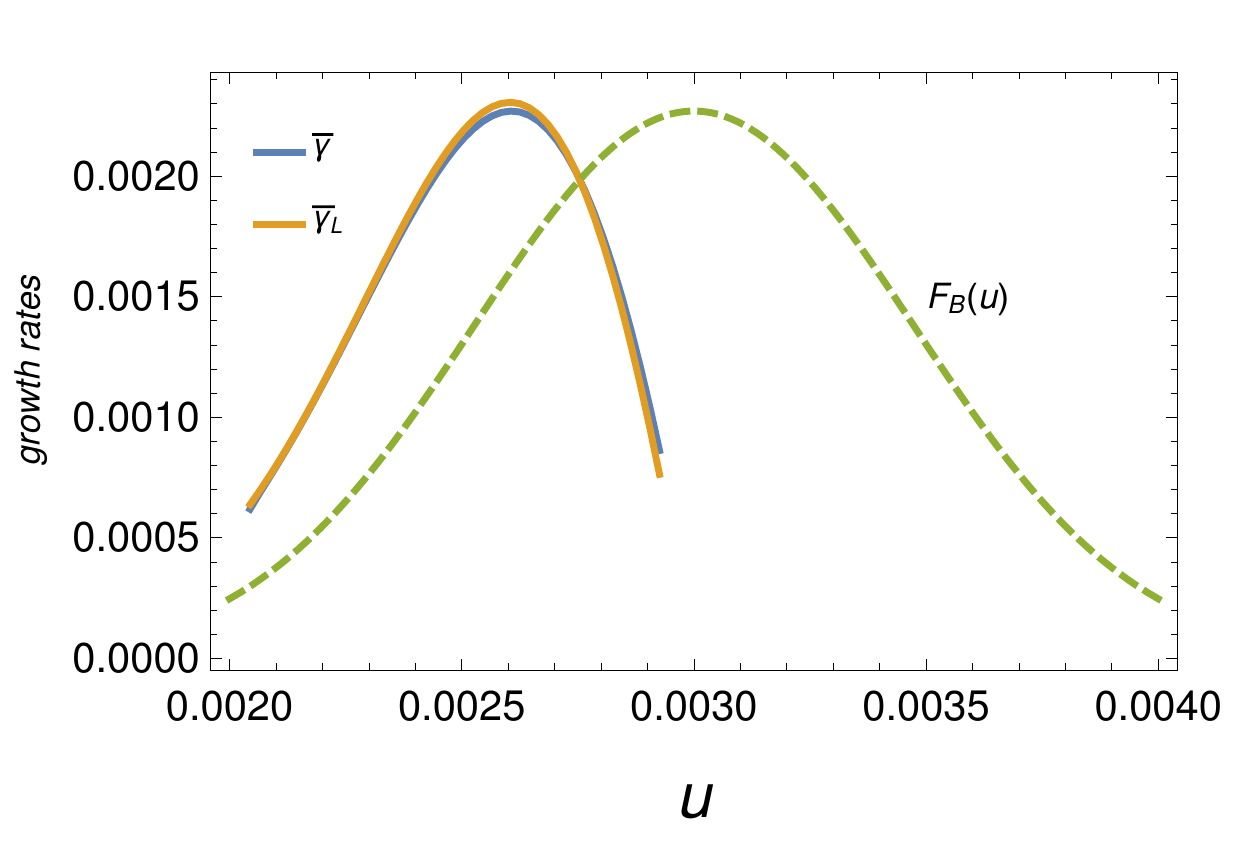}
\includegraphics[width=.405\textwidth,clip]{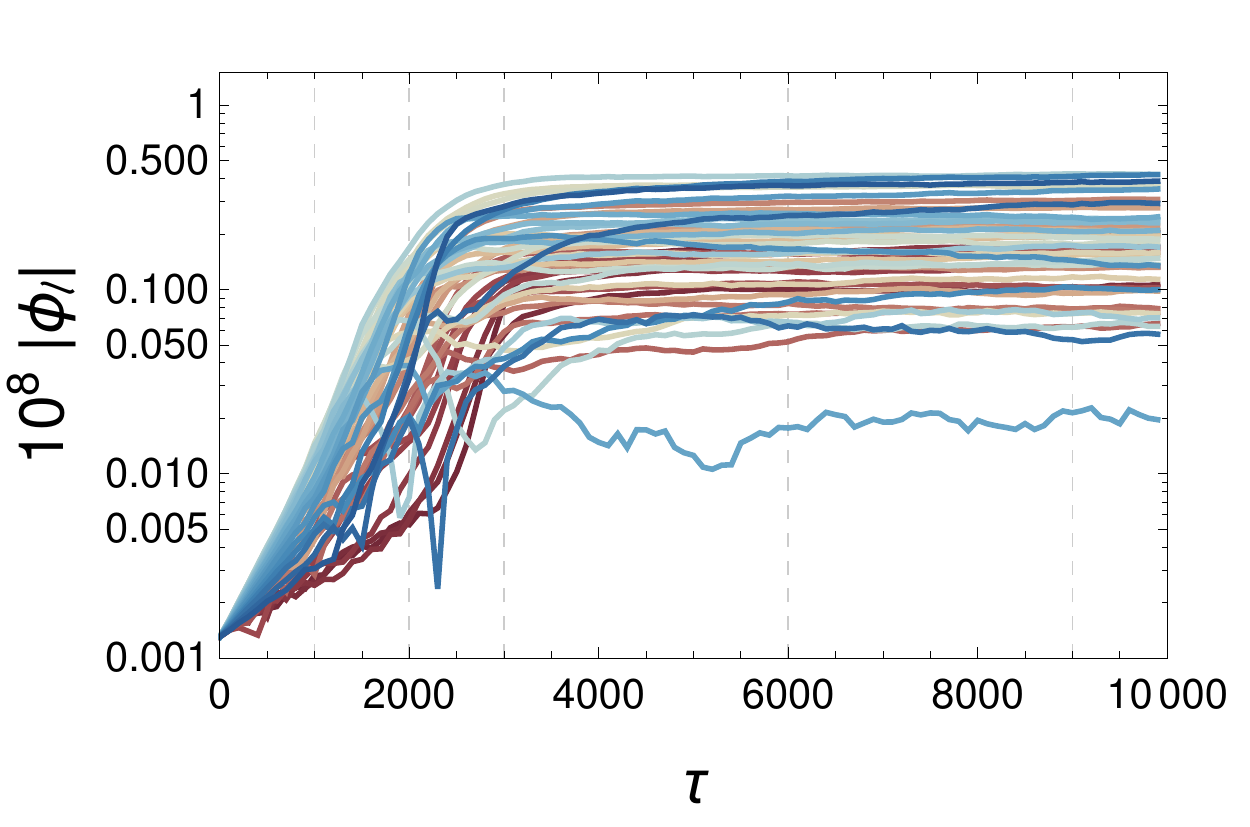}
\caption{(Color online) Left-hand panel: Plot of $\bar{\gamma}$ from \eref{disrel} (blue) and $\bar{\gamma}_L$ from \eref{drlin} (yellow) for several values of the resonant velocity $u_{\ell}$ (in the figure simply denoted by $u$). We also depict (off-scale) the beam initial distribution function $F_B(u)$ (dashed-green) to well represent the resonance positions. Right-hand panel: Temporal evolution of the 45 simulated modes from \erefs{mainsys1} using a standard color scheme from $\ell_{min}=342$ (red) and $\ell_{max}=489$ (blue). Dashed lines denote relevant instants of time taken into account in the next figures: $\tau=[1000,\,2000,\,3000,\,6000,\,9000]$.
\label{fig1}}
\end{figure}

\subsection{Numerical results}
We assume an initial Gaussian warm beam distribution function in velocity space $F_B(u)=\text{Exp}[a(b+u)^{2}]$ (with $a\simeq-2.2\times10^{-6}$ and $b=-0.003$). Having set $\eta=0.0002$, we select and simulate modes for which the relative error between $\bar{\gamma}$ (numerical integration of \eref{disrel}) and $\bar{\gamma}_L$ (\eref{drlin}) is less than $5\%$ (see left-hand panel of \figref{fig1}, where we plot the corresponding values and we also represent (off-scale) the intial distribution function to underline the mode positions). In the non-linear simulations of \erefs{mainsys1}, we run 45 modes with equispaced resonant velocities and $\ell_{min}=342$ and $\ell_{max}=489$, thus getting $\bar{\mathcal{N}}\simeq0.3$.

The mode temporal evolution is plotted in the right-hand side of \figref{fig1}, where the linear and non-linear phase of the dynamics can be easily recognized. We underline how the mode behavior depicted in \figref{fig1} is consistent with the emergent QL approximation in the sense described in Sec.\ref{sec2} (see also \citer{ncentropy}). In fact, the present setup corresponds to an averaged estimate $\tau_{ac}/\tau_{B}\simeq0.02$.
\begin{figure}[ht!]
\centering
\includegraphics[width=.4\textwidth,clip]{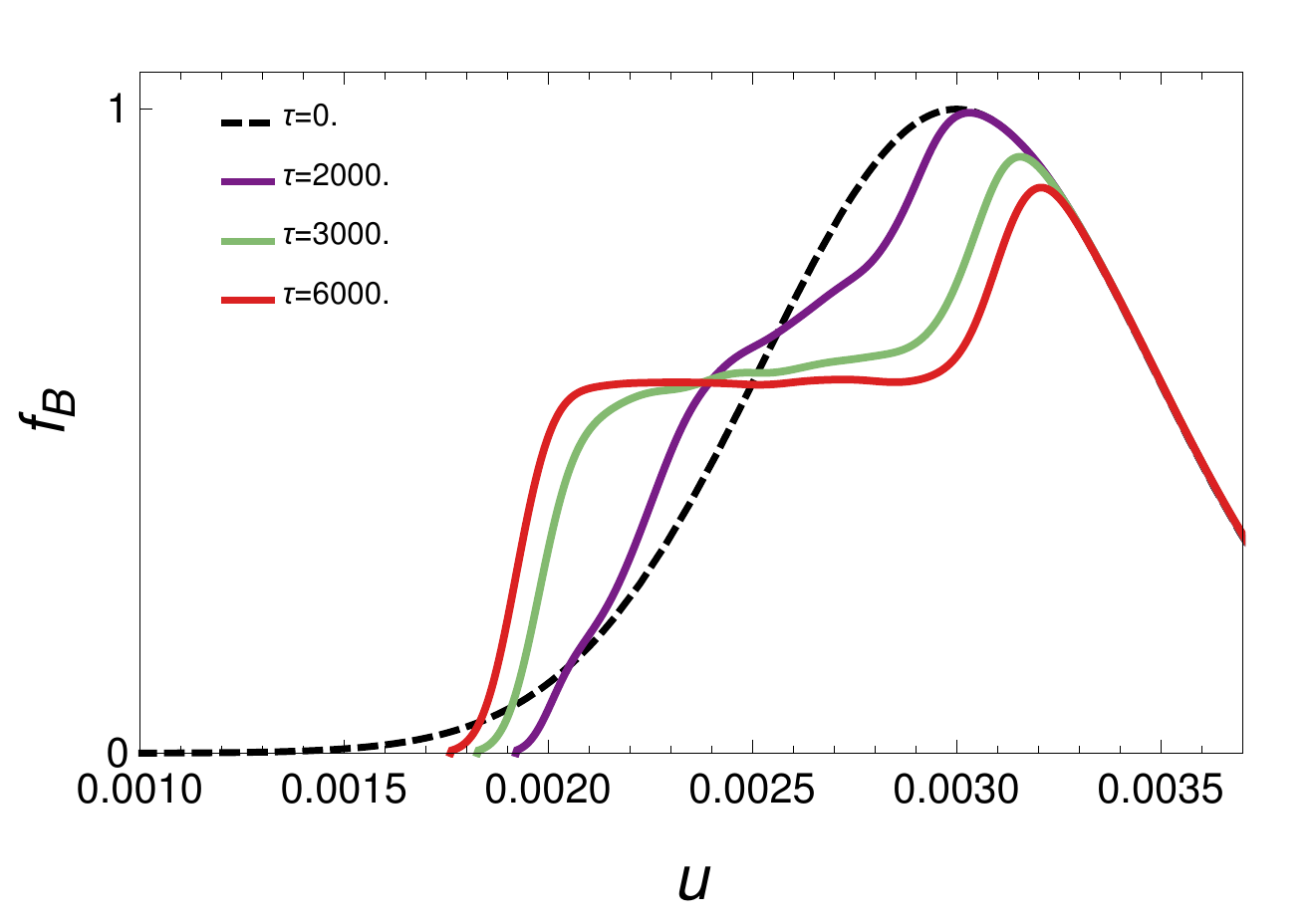}
\includegraphics[width=.4\textwidth,clip]{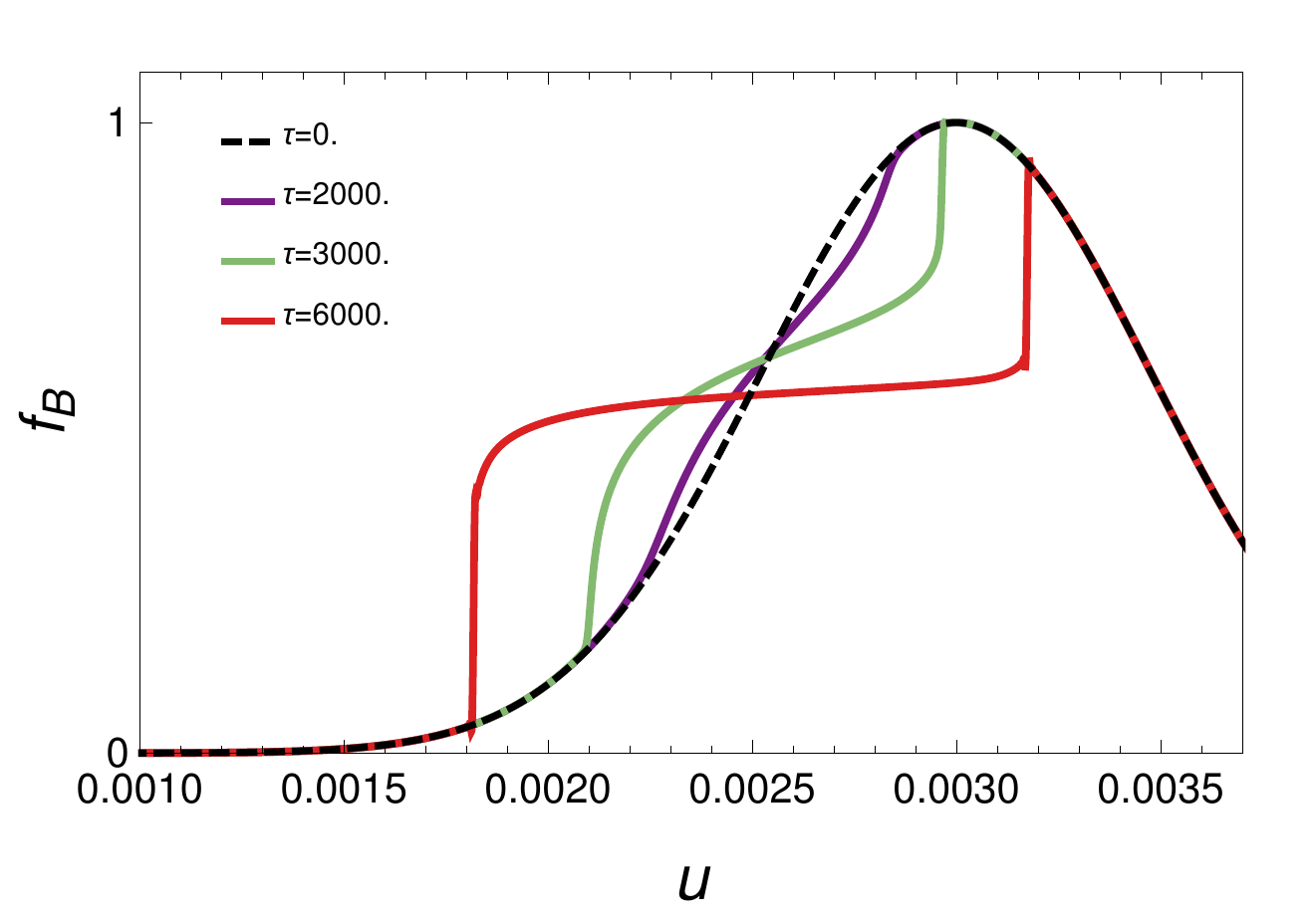}
\caption{(Color online) Evolution of the beam distribution function $f_B(\tau,u$) derived from simulations of \erefs{mainsys1} (left-hand panel) and from the integration of \erefs{QL_u}-\reff{spect} (right-hand panel), taken at given different times as indicated in the plot.
\label{fig2}}
\end{figure}
\begin{figure}[ht!]
\centering
\includegraphics[width=.32\textwidth,clip]{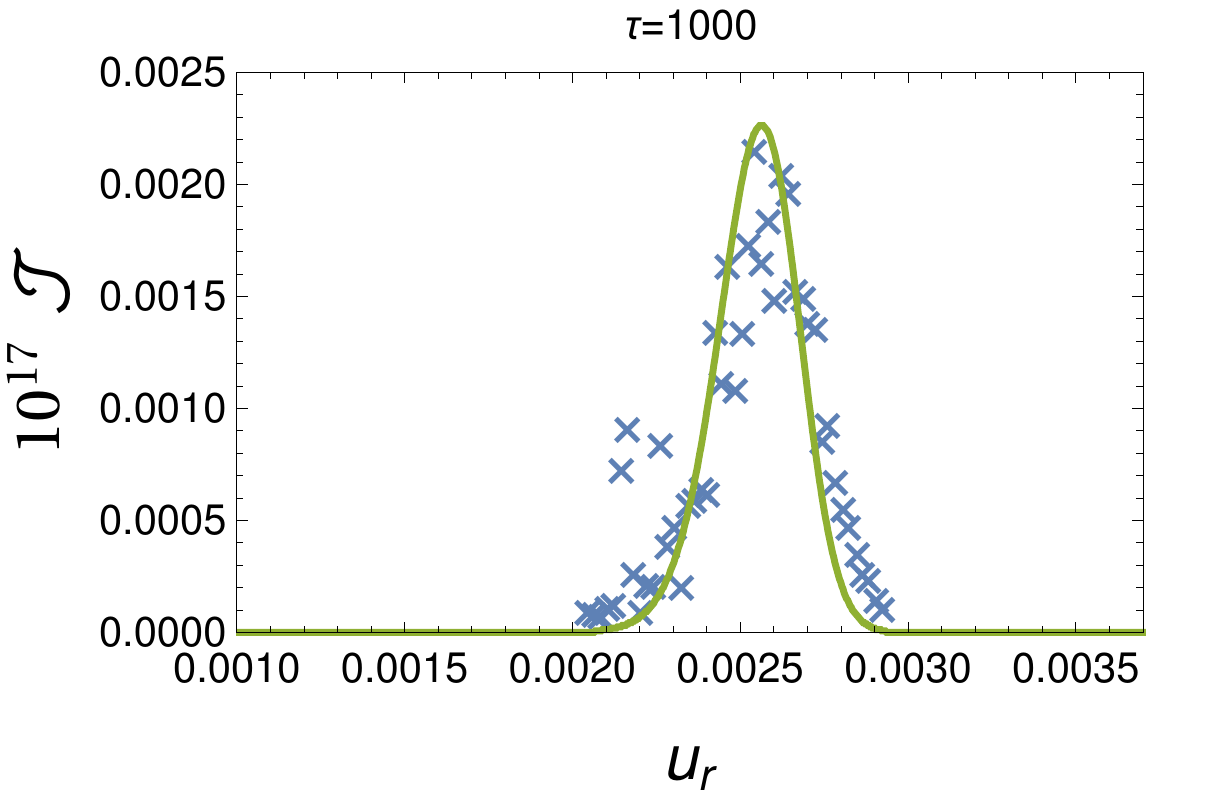}
\includegraphics[width=.3\textwidth,clip]{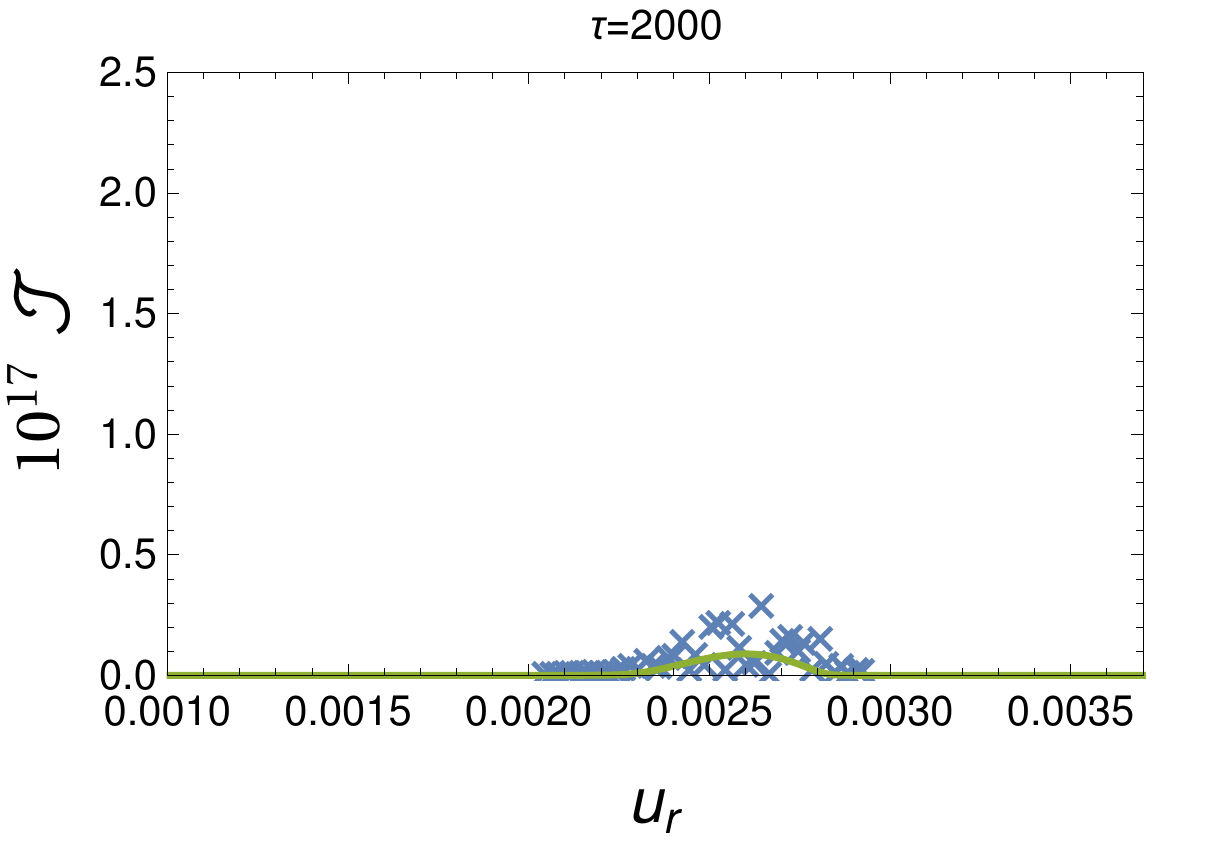}
\includegraphics[width=.3\textwidth,clip]{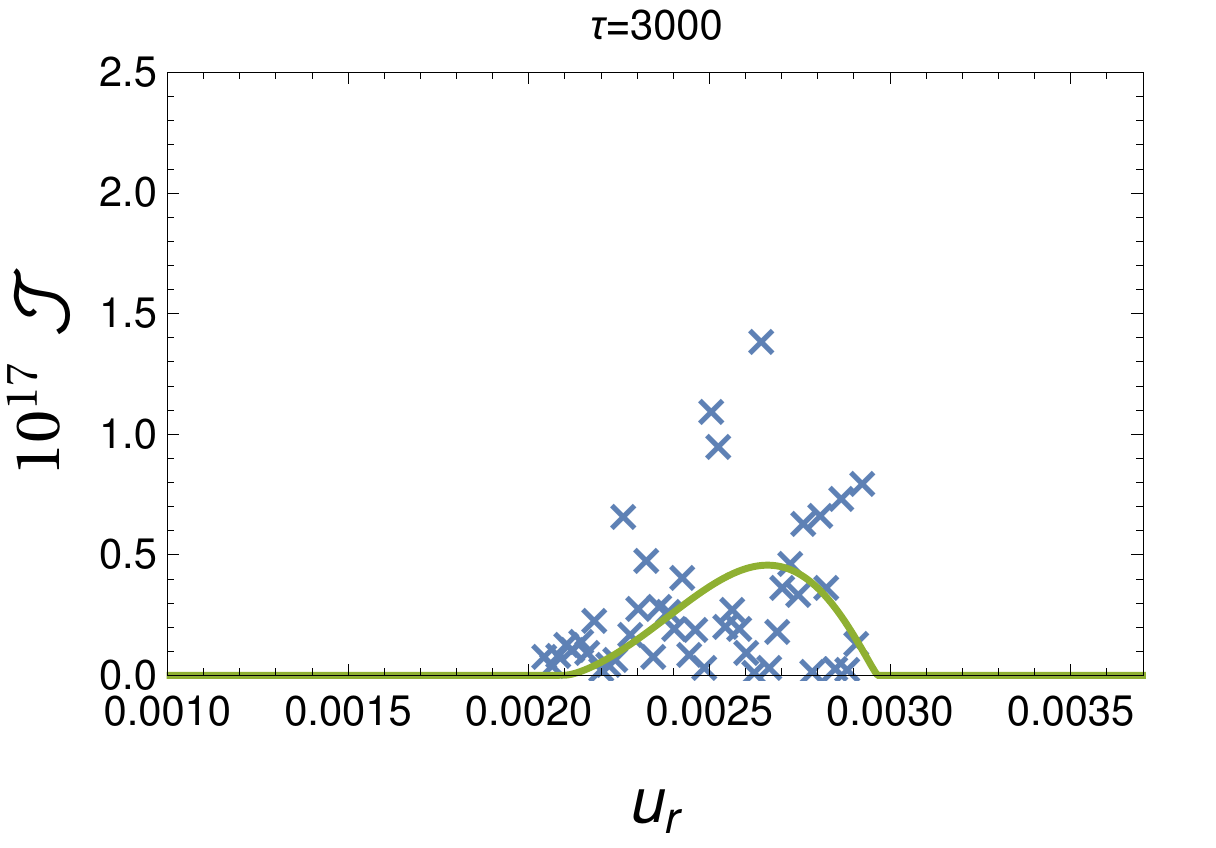}\\
\includegraphics[width=.3\textwidth,clip]{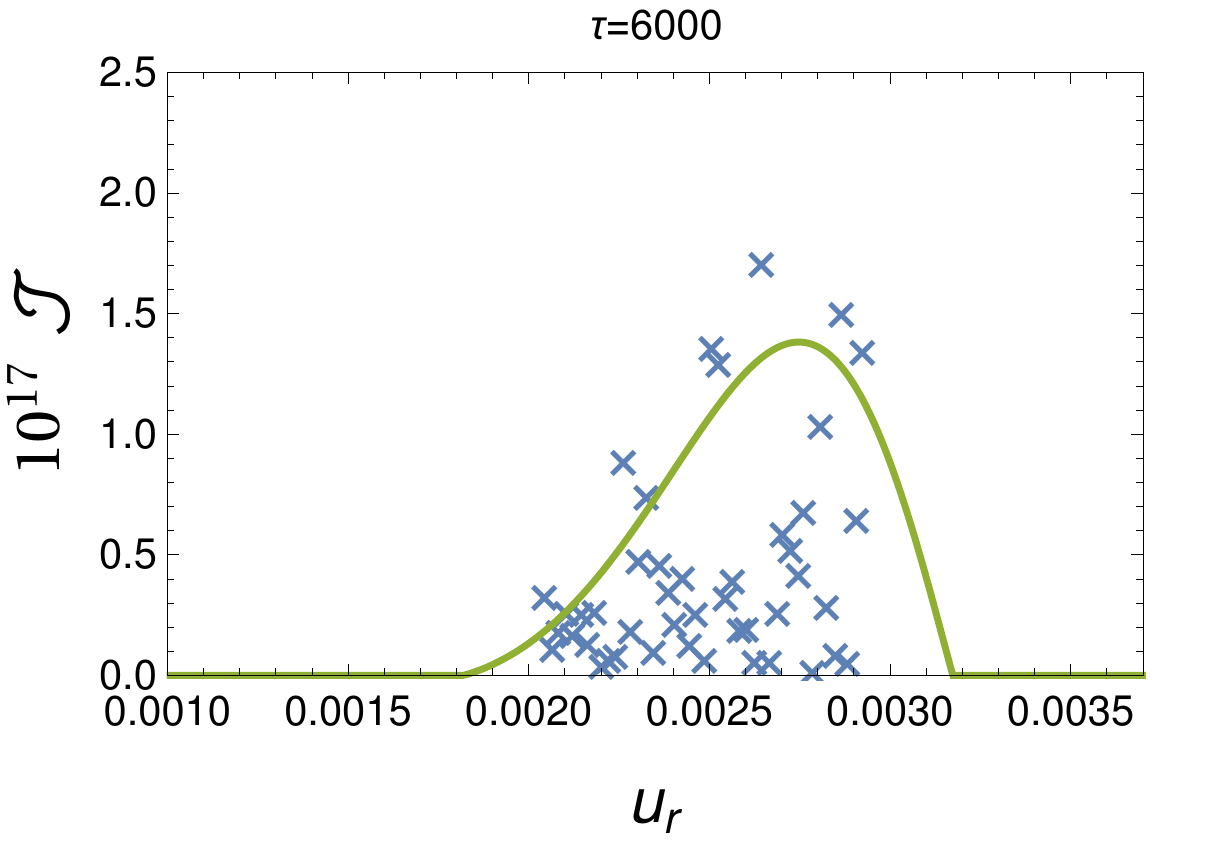}
\includegraphics[width=.3\textwidth,clip]{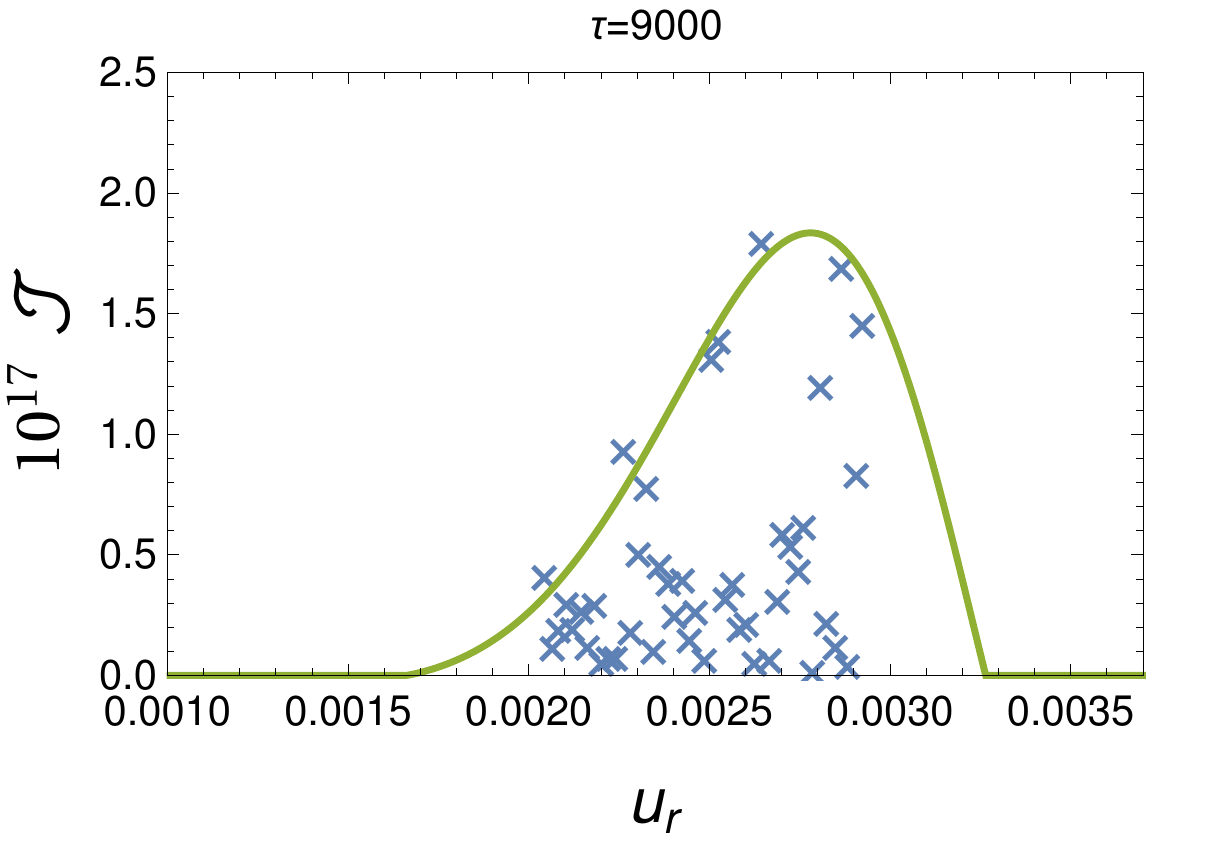}
\caption{(Color online) Plot, at different times (as denoted over the figures), of $\mathcal{J}(\tau,u)$ from \erefs{spect} (green-solid line) and the values of $|\phi_{\ell}|^{2}$ for each $\ell$ (blue crosses) from simulations of \erefs{mainsys1}.
\label{fig3}}
\end{figure}

Let us now compare the evolution of the distribution function as derived from simulations (left-hand panel of \figref{fig2}) with respect to the numerical integration of \erefs{QL_u}-\reff{spect} (right-hand panel of \figref{fig2}). The single PDE for the spectral evolution (\eref{spect}) is integrated using a (linearized) Crank-Nicolson algorithm. The initial condition for $\mathcal{J}(0,u)$ is set Gaussian with features (amplitude, mean and variance) chosen to mimic the discrete simulated mode spectrum.

The comparison of the two behaviors clearly shows a time delay in the evolution of the QL model with respect to $N$-body simulations. In fact, from \figref{fig2} we see that the formation of the plateau (green curve) is slower in the QL model with respect to simulations. This result is consistent with the assumptions under which we derived the QL equations. By other words, the ansatz that the distribution function evolves slowly enough to be considered as constant in \eref{xcxx} affects the whole evolution of the QL model, making it less reactive in time than the real system. 

Analogously, we can analyze the behavior of the spectrum in the two schemes. In \figref{fig3}, we over-plot (for different times) the function $\mathcal{J}$ integrated from \eref{spect} and the values of $|\phi_{\ell}|^{2}$ extracted from simulations. We see again that the QL spectrum evolves more slowly since $\tau\simeq1000$ and it is just this behavior which is responsible for the discrepancy in the distribution function according to \eref{QL_u}. Nonetheless, at sufficiently late times, the QL spectrum satisfactory envelops the simulation results \footnote{We would like to underline that a certain degree of scattering in the numerical spectrum determined via the $N$-body simulations is not depending on the number of simulated particles, but it is associated to the limit our mode number to reproduce the real continuum of the Langmuir spectrum. However, the addressed modes are adequate to reproduce all the feature of a QL profile and the resulting spectral morphology does not prevent a satisfactory comparison with the PDE numerical plots.}. It is worth noting that \eref{spect} naturally contains the linear regime of the instability. In fact, when the field intensity is very small (according to the level of the Langmuir fluctuations), we naturally recover the linear growth of the intensity according to the standard inverse Landau damping rate. However, as soon as the non-linear contribution grows enough, it accounts for a diffusive-like behavior in the velocity space, which has no counterpart in the real simulations where such effects are absent.

Since, as discussed above, we set up our initial conditions in order to avoid non-perturbative effects in the dispersion relation, we can claim that, in the initial linear phases, the numerical and QL spectrum have to coincide, as clearly emerges from the first panel of \figref{fig3} ($\tau=1000$). The crucial point is that the diffusive contributions, naturally associated with the Fokker-Planck equation \reff{st1}, induces a too early termination of the linear regime with respect to the exact simulations. As a consequence, the spectrum remains at a slower level with respect to the real one and the whole temporal mesoscale evolution turns out to be delayed. These considerations suggest the necessity of a revised approach able to account for the coherent particle-field interactions in the mesoscale temporal evolution of the BPS.

Finally, it is important to stress that the different width of the plateau region in the two approaches (in particular, cf. the red curves of \figref{fig2}) has to be mainly attribute to the fact that the QL spectrum is larger than the numerical one according to the presence of a continuum of linearly stable modes, against the fixed character of the mode array used in the simulations.

\section{Corrections to the instantaneous QL growth rate}\label{sec7}
The comparison with simulations outlines how the system reactivity at mesoscales is underestimated in the QL model. We are now going to show how a more reactive system is predicted by relaxing the quasi-stationary assumption for the distribution function. In Fourier space, the Vlasov-Poisson coupled system writes as
\begin{align} 
&\p_t f_k(t) = -ikvf_k + \frac{e}{m}\sum _q E_{k-q}\partial_vf_q\;,\label{a2} \\
&\p_t E_k(t) = - i \omega _p E_k(t) + \frac{2\pi e\omega _p}{k}\int_{-\infty}^{+\infty}\!\!\!\!\!\!\! f_k(t,v) dv \;.
\label{pre11}
\end{align}
%Using the homogeneity condition \eref{ql6},
We now implement the assumption that the $k$ component of the distribution function receives mainly contribution just by the correspondent harmonics. Setting $q=0$ in \eref{a2}, we get the solution
\begin{equation} 
f_k(t,v) = \frac{e}{m} \int_0^t dt^{\prime} E_k(t^{\prime}) \exp \big[ikv(t^{\prime}-t)\big]\p_v f_0(t^{\prime},v)\;.
\label{sm} 
\end{equation}
Without loss of generality, we can set the electric field as
\begin{equation}
E_k(t)=\bar{E}_k \exp \Big[ -i\int_0^t \omega_k(t^{\prime})dt^{\prime} \Big]\;,\qquad
E_k(t^{\prime})=E_k(t)\exp \Big[-i\int_t^{t^{\prime}}\omega_k(y)dy\Big]\;,
\label{pre5}
\end{equation}
where $\bar{E}_k=const.$, $\omega_k=\omega_p+\delta \omega_k$, while $\delta\omega_k$ is a small complex function. Using the last expression in \eref{sm}, the dynamics of $f_0$ assumes the following Dyson-like form
\begin{align} 
\p_t f_0(t,v) = \frac{e^2}{m^2}\sum_{k>0} |E_k(t)|^2
\p_v \Big(&\int_0^tdt^{\prime}\exp\Big[ikv(t^{\prime} - t) - i\int ^{t^{\prime}}_tdy \,\omega _k(y)\Big]\p_vf_0(t^{\prime},v)+ c.c.\Big)\;,
\label{so}
\end{align}
where the reality of the electric field allowed to use the relations $E_{-k} = E^*_k$ and $\delta\omega_{-k}=-\delta\omega^*_k$. For a derivation of an equivalent Dyson-like equation in the $\omega$-space, see \citers{AK66,ZCrmp}.

Similarly, \eref{pre11} can be manipulated leading to the following evolutive equation for the electric harmonics intensity
\begin{align}
\p_t |E_k|^2 = \frac{2\pi e^2\omega _p}{mk} |E_k|^2
\Big( \int_{-\infty}^{\infty}dv\int_0^tdt^{\prime}
\exp\Big[ikv(t^{\prime} - t) - i\int ^{t^{\prime}}_tdy \,
\omega _k(y)\Big]\partial _vf_0(t^{\prime},v)+c.c.\Big)\;.
\label{sq}
\end{align}
The system of \erefs{so} and \reff{sq} constitutes the basic framework to self-consistently evolve the beam-plasma interaction under the hypothesis of average homogeneity only.

If in the region where the distribution function $f_0$ remains significantly different from zero (namely $v_{min}<v<v_{max}$), we assume that the Langmuir spectrum is sufficiently dense in order to have $kv\simeq \omega_p$ for $\forall v$, the integral taken in $dt'$ of the two expressions above has a small exponent (the contribution $(t^{\prime} - t)\omega_p$ cancels out between the two parts of the exponent). In this case, such an integral can be rewritten as 
\begin{equation}
\int_0^tdt^{\prime} \frac{1}{i(kv - \omega_k(t^{\prime}))} \partial _vf_0(t^{\prime}, v) 
\partial_{t^{\prime}}\Big( 
\exp\Big[ikv(t^{\prime} - t) -  
i\int_t^{t^{\prime}} 
\omega_k(y)dy\Big]\Big)\;. 
\label{xcxx}
\end{equation}
If $\omega_k$ is not a fast varying function in time and we Taylor expand $f_0$, retaining only the next-to-the-leading order term, {\it i.e.} $\partial_v f_0(t',v)=\partial_v f_0(t,v)+(t'-t)\,\partial_t\partial_vf_0(t,v)$, the expression \reff{xcxx} rewrites  
\begin{align}
\int_0^tdt^{\prime} \frac{1}{i(kv - \omega _k)} \partial _vf_0(t, v) 
	\partial _{t^{\prime}}\Big( 
	\exp\Big[ ikv(t^{\prime} - t) - 
		i\int_t^{t^{\prime}} 
\omega_k(u)du \Big]\Big) +\qquad\quad\nonumber\\
-\int_0^tdt^{\prime} \frac{1}{i(kv - \omega _k)} \partial_t\partial _vf_0(t, v)  
	\exp \Big[ ikv(t^{\prime} - t) - 
		i\int_t^{t^{\prime}} 
\omega_k(u)du \Big]\,,
\label{xcxx2}
\end{align}
and the integrals can be explicitly computed (see appendix A in \citer{OWM71}) finding
\begin{equation}
\frac{1}{i(kv - \omega _k(t))}\p_vf_0(t,v)+\frac{1}{(kv - \omega _k(t))^2}\p_t\p_vf_0(t,v)\;. 
\label{ss} 
\end{equation}
A similar Taylor expansion is performed in \citer{Klimontovich}, where it 
is applied to the calculation of the collisional integral, from which the 
spectral densities of the particle numbers and of the electric field are evaluated
for non-ideal stationary and homogeneous plasmas, where the 
fluctuations are non-stationary and the delayed contribution could 
be relevant.

Finally, \eref{so} can be restated as as 
\begin{align}
&\p_t f_0 = \p_v(\mathcal{D}_{QL}(t,v)\p_v f_0+\mathcal{D}_1(t,v)\partial_t\partial _vf_0)\;,\label{st1X}\\
&\mathcal{D}_{QL}(t,v) \equiv 
\frac{e^2}{m^2}\sum _{k>0}
|E_k(t)|^2\Big[\frac{1}{ 
i(kv - \omega _k)}-\frac{1}{ 
i(kv - \omega^* _k)}\Big]\;. 
\label{st2X} \\
& \mathcal{D}_1(t,v) \equiv 
\frac{e^2}{m^2}\sum _{k>0}
\mid E_k(t)\mid^2\Big[\frac{1}{ 
\left(kv - \omega _k\right)^2}+\frac{1}{ 
\left(kv - \omega^*_k \right)^2}\Big]\; . 
\label{D1} 
\end{align}
As already discussed, the QL diffusion coefficient $\mathcal{D}_{QL}$ in \eref{st2X} is found by approximating the discrete $k$-space by a continuum, \ie $\sum _k\left(...\right) \to\int _{k_{min}}^{k_{\max}}dk \;\mathcal{N}(k)(...)$ and assuming a constant state density $\mathcal{N}$, thus getting the formal expression in \eref{st2}.
%\begin{align}
%\mathcal{D}_{QL}(t,v) \sim \frac{e^2\pi\mathcal{N}}{2m^2v}\Big[ 
%|E(t,\omega _k/v)|^2+|E(t,\omega^* _k/v)| ^2\Big]
%\simeq \frac{e^2\pi\mathcal{N}}{m^2v}| E(t,\omega_p/v)| ^2\;,
%\label{sb2}
%\end{align} 
It is worth stressing how, in the QL model, the assumption that the function $f_0$ is quasi-stationary relies on the idea that the velocity beam profile reaches a stationary regime in the resonance region.

Applying similar considerations to the time integral in \eref{sq}, we get the following generalized expression of \eref{sv}: 
\begin{align}
&\p_t|E_k(t)|^2 = 2\Gamma_k |E_k|^2\;,\label{svX}\\
&\Gamma_k(t)=\gamma^{QL}_{k}+\delta\gamma_k\;,\\
%&\gamma^{QL}_{k}(t) = -i\frac{\pi e^2\omega_p}
%{mk}\int_{-\infty}^{\infty}dv\Big[  
%\frac{1}{kv - \omega _k} 
%- \frac{1}{kv - \omega_k^*}
%\Big]\partial_v f_0\;,\\
&\delta\gamma_k(t)=\frac{\pi e^2\omega_p}{mk}
\int_{-\infty}^{+\infty}dv\Big[  
\frac{1}{\left(kv - \omega _k\right)^2} 
+ \frac{1}{\left(kv - \omega_k^*\right)^2}
\Big]\partial_t\partial _vf_0\;,
\label{svbis}
\end{align}
where we mention that the QL growth rate $\gamma^{QL}_{k}$ (see \eref{svGamma}) is obtained for $\mathrm{Im}(\omega_k)\ll\omp$. The additional contribution $\delta\gamma_k$ has already been written in \citer{D82} (see Eq.(49) in chapter 14), where it is called the non-resonant contribution to the growth rate, but it is not evaluated and simply neglected with respect to the non-linear wave-particle and wave-vave interactions. Its impact on the BPS is here addressed.
 
We can rewrite the integral in \eref{svbis} via a partial integration as follows
\begin{equation}
2\,\int_{-\infty}^{+\infty}dv
\frac{(kv-\omega_p)^2-\Gamma_k^2}{\left[(kv - \omega _p)^2+\Gamma_k^2\right]^2}\,\partial_t\partial _vf_0
=-2\,\int_{-\infty}^{+\infty}dv \,G(kv - \omega _p)\, \partial_tf_0\;,
\end{equation}
where the function $G(y)$ within the integral reads 
\begin{equation}
G(y)=k\,\partial_y\Big[\frac{y^2-\Gamma_k^2}{\left[y^2+\Gamma_k^2\right]^2}\Big]=2k\,y\,\frac{3\Gamma_k^2-y^2}{\left[y^2+\Gamma_k^2\right]^3}\;.
\end{equation}
This function has absolute maximum and minimum at $y=\pm (\sqrt{2}-1)\,\Gamma_k=\pm \bar{y}$ and it rapidly tends to vanish for $y\gg\Gamma_k$. Hence, we split the integral into the sum of the two contributions for negative and positive $y$ and we approximate the function $\partial_v f_0$ by its values at $\pm\bar{y}$, so getting\footnote{Here, we approximate the variation of the function $\partial_tf_0$ by its derivative with respect to velocity, valid for $\Gamma_k\ll \omega_p$.}  
\begin{align}
-2\,\int_{-\infty}^{+\infty}dv \, G(kv - \omega _p)\, \partial_tf_0\sim 
-2\partial_tf_0(kv=\omega_p-\bar{y})\int_{-\infty}^{\omega_p/k}dv\, G(kv - \omega _p)+\qquad\qquad\nonumber\\
-2\partial_tf_0(kv=\omega_p+\bar{y})\int^{+\infty}_{\omega_p/k}dv\, G(kv - \omega _p)\sim-\frac{4(\sqrt{2}-1)}{k\Gamma_k}
\partial_v\partial_t f_0|_{v=\omega_p/k}\;.
\end{align}
Therefore, $\delta\gamma_k$ reads
\begin{equation}
\delta\gamma_k= -\frac{4(\sqrt{2}-1)\pi e^2\omega_p}{mk^2\Gamma_{k}}\, \partial_t\partial_vf_0|_{v=\omega_p/k} \;,
\end{equation}
and the following equation can be written for $\Gamma_k$
\begin{equation}
\Gamma_k=\gamma^{QL}_{k} -\frac{4(\sqrt{2}-1)\pi e^2\omega_p}{mk^2\Gamma_{k}}\, \partial_t\partial_vf_0|_{v=\omega_p/k}\;.
\end{equation}

Using the dimensionless variables introduced for \erefs{QL_u} and \reff{spect}, the solution of the equation above writes as
\begin{align}\label{gamma_}
\bar{\Gamma}=\frac{\bar{\gamma}^{QL}}{2}+\frac{\bar{\gamma}^{QL}}{2} \sqrt{1-4\bar{\delta}^{2}/(\bar{\gamma}^{QL})^{2}}\;,\quad \textrm{with}\quad
\bar{\gamma}^{QL}(u)=\frac{\pi \eta u^{2}}{2 M}\,\p_u f_B\;,\quad
\bar{\delta}^{2}(u)=\frac{(\sqrt{2}-1) \eta u^{2}}{M}\,\p_\tau\p_u f_B\;,
\end{align}
where we recall that barred growth rates are expressed in $\omp$ units and it is worth noting how a small deviation from the QL growth rate is found for $4|\bar{\delta}^2|\ll (\bar{\gamma}^{QL})^2$. The predicted correction to the QL growth rate is expected to be positive approaching the plateau region, where the distribution function flattens and $\partial_u f_B$ is a decreasing function of time, giving a negative value for $\bar{\delta}$. As a consequence, the value of the modified growth rate is larger than the QL instantaneous one. 

The modified expression for the linear instantaneous growth rate could allow us to restate also the equation for the distribution function into an upgraded form able to account for the mesoscale dynamics. Up to first approximation, we could infer that this non-steady correction affects the diffusion coefficient only, leaving unchanged the morphology of \eref{st1} accordingly with the considerations made in \citer{Laval99} about the proper renormalization of the theory.

It is worth stressing that our correction to the QL theory is valid as far as $(t'-t)$ remains a small quantity in the Taylor expansion of the distribution function, and this can take place for sufficiently early times only (surely it is not true in the late phases). In this respect, we can get an analytical form of the ratio between the revised expression for the spectral intensity and the QL one, at least when, during the early mesoscale evolution, they do not differ too much, \ie when $\bar{\delta}^{2}/(\bar{\gamma}^{QL})^{2}\ll1$. From \eref{svX}, we naturally obtain the relation
\begin{equation}
\mathcal{J} = \mathcal{J}(0,u)\exp\Big[\int _0^\tau 2\bar{\Gamma}(\tau',u)d\tau^{\prime}\Big]\;,
\label{sz}
\end{equation}
and, in the considered limit, we can perform the following expansion of \eref{gamma_}:
\begin{align}
\bar{\Gamma}=\bar{\gamma}^{QL}-\bar{\delta}^2/\bar{\gamma}^{QL}\;.
\end{align}
Substituting this expression in \eref{sz} and taking the primitive of the exact differentiation, it naturally provides the relevant form of the ratio $\mathcal{J}/\mathcal{J}_{QL}$, \ie
\begin{align}\label{bnhs}
\frac{\mathcal{J}}{\mathcal{J}_{QL}}= \Big(\frac{|\p_u F_B|}{|\p_u f_B|}\Big)^{\pi(\sqrt{2}-1)}\;,
\end{align}
where, henceforth, $\mathcal{J}_{QL}$ denotes the QL spectrum explicitly plotted in \figref{fig3}.

Since the derivative of the distribution function clearly decreases with respect to the Gaussian initial condition as the system evolves, it naturally comes out that the expression above is, on average, describing an enhancing of the predicted spectrum with respect to the QL one.
\begin{figure}[ht!]
\centering
\includegraphics[width=.42\textwidth,clip]{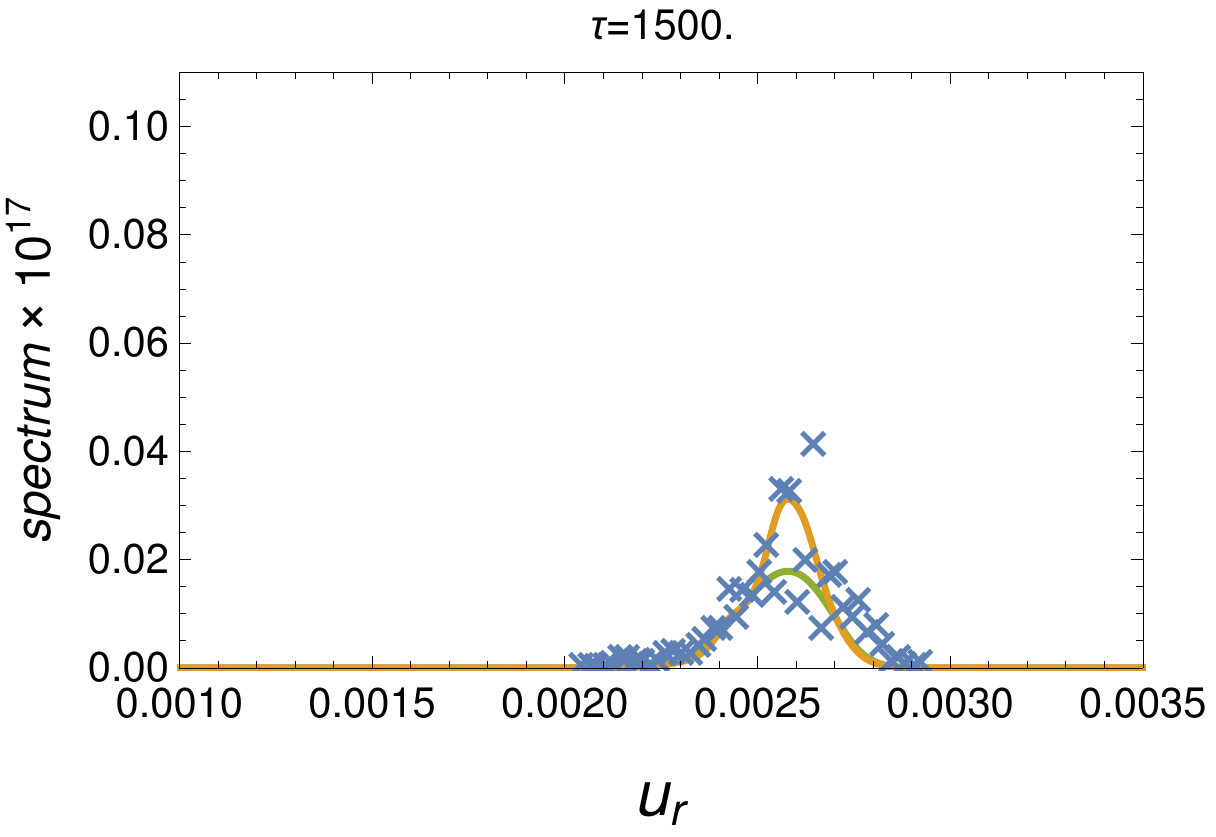}
\includegraphics[width=.42\textwidth,clip]{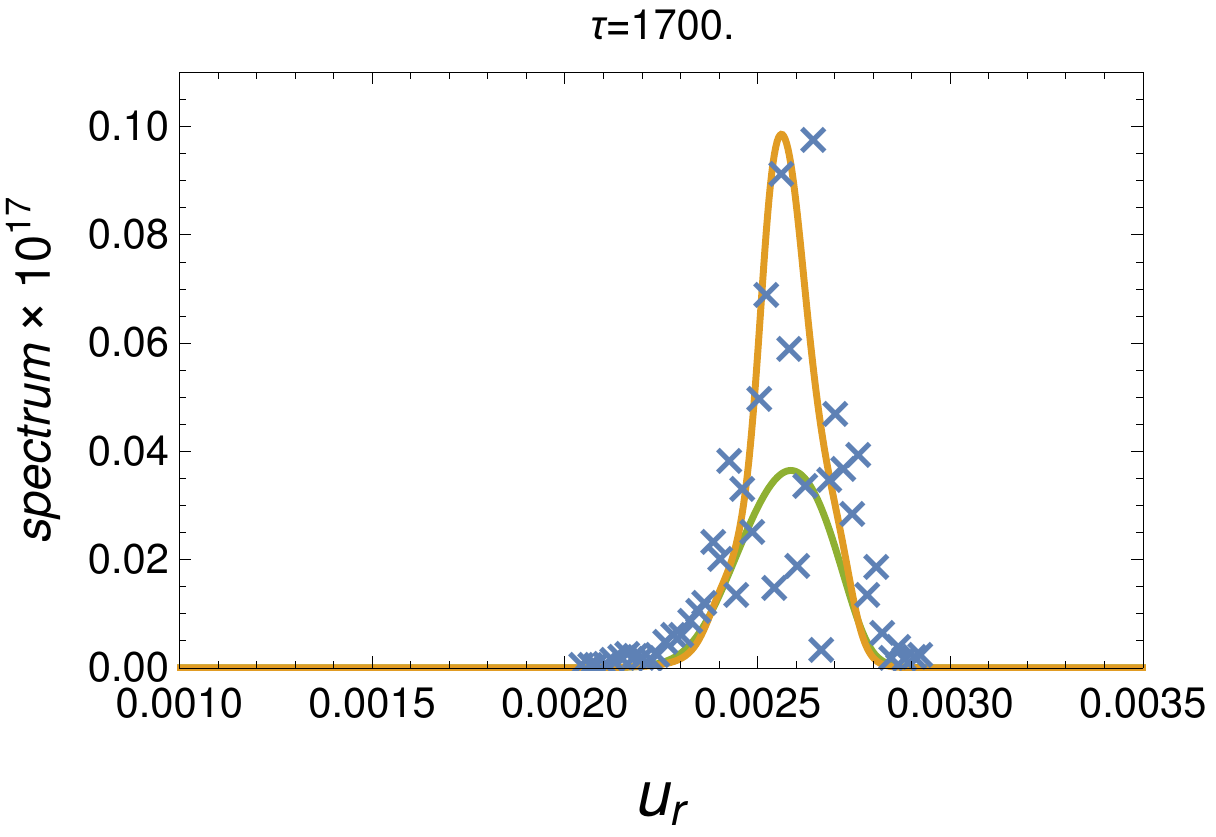}
\caption{(Color online) Plot of $\mathcal{J}_{QL}$ (green, cf. \figref{fig3}), $|\phi_{\ell}|^{2}$ for each $\ell$ (blue crosses) from simulations and $\mathcal{J}$ (orange) from \eref{bnhs}. Also the profiles $f_B(u,\tau)$ are extracted from the simulation results. \label{fig_spcfr_}}
\end{figure}
As a check, we use, for the early temporal mesoscales of the system (says $\tau\simeq1600 $) the numerically calculated distribution function to 
estimate the ratio above. By other words, we use the distribution function extracted from the $N$-body dynamics to calculate $\p_u f_B$ at the considered times. As shown in \figref{fig_spcfr_}, the amended spectrum (orange solid line) is significantly close to the numerical one (we also plot the original QL spectrum for comparison). This good agreement between the 
predicted and the simulated spectra indicates the 
validity of our derivation for the growth rate and 
that the addressed temporal stages are consistent 
with a small deviation from the QL model, when \eref{bnhs} reliably holds.

In order better understand the dynamical implications of the amended dynamics, we now evaluate the ratio $\bar{\Gamma}/\bar{\gamma}^{QL}$ from \eref{gamma_} in the full mesoscale temporal region, \ie when \eref{bnhs} is no longer valid. 
\begin{figure}[ht!]
\centering
\includegraphics[width=.42\textwidth,clip]{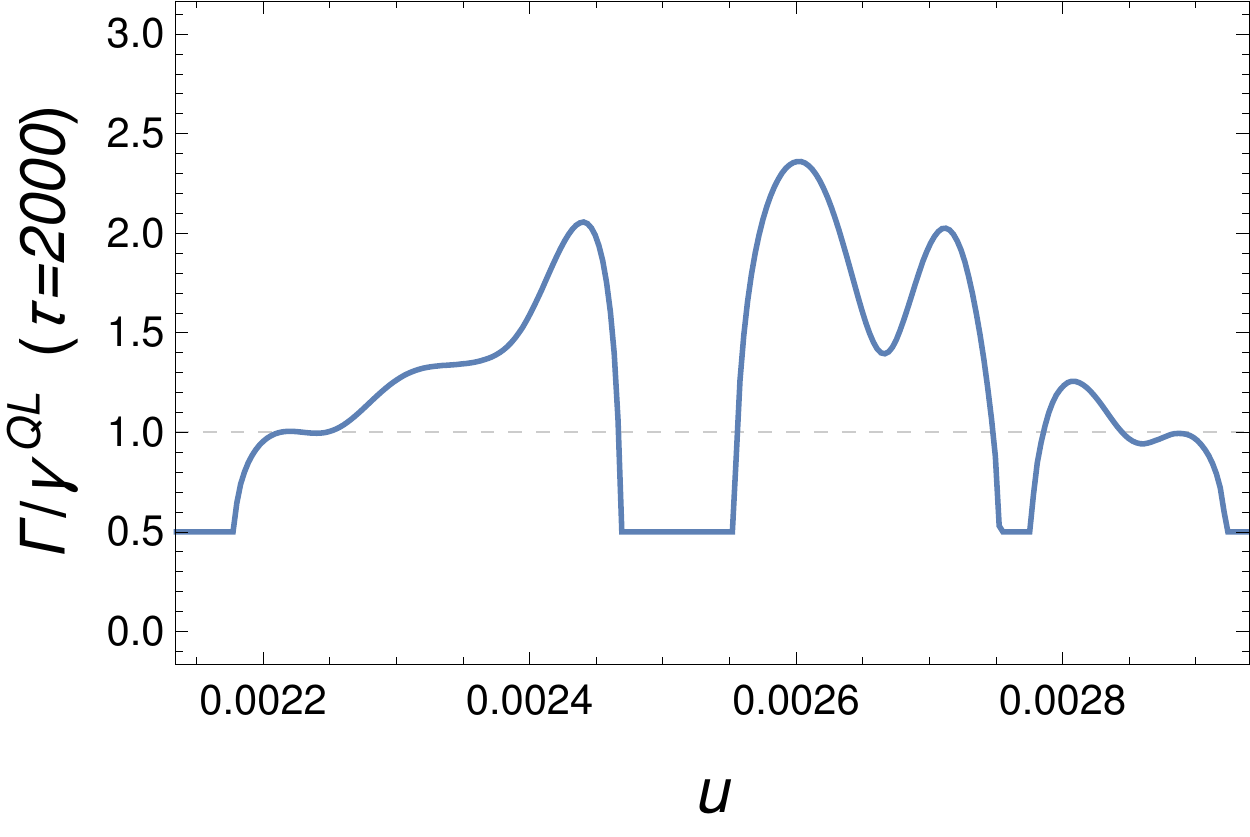}
\includegraphics[width=.42\textwidth,clip]{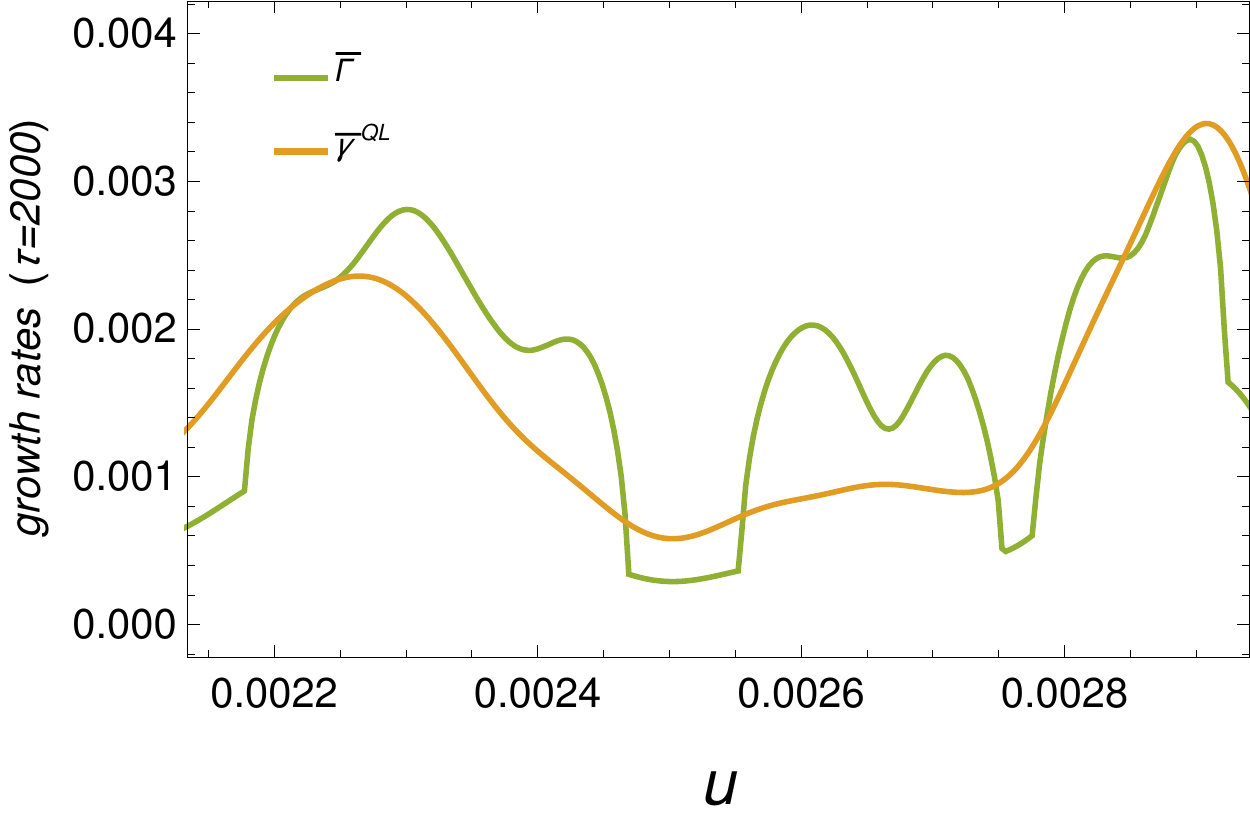}
\caption{(Color online) Left-hand panel: Plot of the real part of the ratio $\bar{\Gamma}/\bar{\gamma}^{QL}$ from \eref{gamma_} as function of the dimensionless velocity for $\tau=2000$. The quantities in \eref{gamma_} are evaluated using the distribution function $f_B(u)$ extracted from the $N$-body simulation. Right-hand panel: Plot of the individual growth rates for the same case. The plots are restricted to the the resonant region addressed in \figref{fig1} (left-hand panel).
\label{fig_gamma_}}
\end{figure}
In particular, we use the morphology of the distribution function at $\tau=2000$, \ie when most of the spectrum is saturating, again by means of the numerical data. This ratio (together with the individual growth rates) is plotted in \figref{fig_gamma_}. We see that $\bar{\Gamma}$ is significantly enhanced with respect to the instantaneous QL growth rate clarifying how, as soon as the system leaves the linear phase of the instability, our calculation provides the requested increase of the system reactivity at the ground of the temporal mesoscale inadequacy of the QL model.

The anaysis above clearly otulines how the present reformulation of the growth rate for the BPS is able to account for deviations at the mesoscale evolution. However, it is worth noting that our approach still relies, like the QL one, on the negligibility of cross-coupling terms among spectral components. Actually, new physical features could be enclosed in a revised picture by addressing such non-diagonal effects as, for instance, consireded in \cite{K69} where the so-called trapped-particle instability is investigated. This scenario relies on the presence of a large amplitude electrostatic mode which is able to trigger sidebands in the spectrum. Such kind of physics is surely relevant for the possible deviation from the QL model but, as far as the beam is sufficiently tenuous, the homogeneity of the problem is on average preserved and the diagonal hypothesis remains reliable during the dynamics.

\section{Conclusions}\label{sec8}
In this work, we compared the dynamical evolution of the QL model with respect to the results obtained from a $N$-body simulation, outlining a qualitative but definitely distinguishable delay of the former with respect to the latter. The nature of our analysis identifies the source of such a discrepancy in the too restrictive character of the quasi-stationary assumption for the function. Thus, we suggest that, in order to significantly upgrade of the 1D QL dynamics, it is relevant to make precise the evolution of the distribution function. In this respect, we amended the QL instantaneous growth rate by accounting for the distribution function evolution via a linear temporal expansion, and we provided the implications on the self-consistent dynamics.

Although not all the original derivations do make explicit account for the quasi-stationary behavior of the distribution function, their statistical characterization of the field and particle fluctuations are implicitly underestimating the real time response of the BPS. Thus, it clearly emerges how the QL model is really predictive only for the linear stages and for the late evolution, while the derivation must be refined for the temporal mesoscales, for instance, as in Sec.\ref{sec7}, taking into account corrections due to the distribution function time dependence.

We validated such corrections to the growth rate, obtained by retaining the next-to-leading order in a Taylor expansion, by using the behavior of the distribution function extracted from numerical simulations. We first evaluated an analytical expression for the ratio between the predicted and QL spectra in those early mesoscale region where they are sufficiently close: we showed how a very good agreement exists between the restated spectrum and the numerical one. We also evaluated the ratio between the modified and the QL growth rates when this approximation is no longer valid and the revised expression was obtained two time larger than the QL one. The obtained result suggests to look for a more appropriate partially differential representation of the system evolution at the temporal mesoscale. This requires the possibility to restate the Vlasov equation in a suitable predictive and differential form. The determination of such a reduced model for the Dyson-Poisson system would constitute a significant improvement in the description of BPS.

Summarizing, it is worth noting how the main goal of our analysis consists of having outlined the necessity to revise the expression of the QL instantaneous growth rate in order the model be able to better fulfill $N$-body simulations. Actually, we also traced the proper way how to get a revise formula, by refining the time evolution of the distribution function contribution to the spectrum of the Langmuir waves. 

The QL model satisfactory accounts for the initial linear phase of the beam-plasma instability and for the late time diffusive evolution in the velocity space, but it significantly fails when applied to the temporal mesoscale evolution. This has relevant implications also for the Tokamak physics, when a one-to-one mapping between the velocity space of the BPS and the radial coordinate of a Tokamak equilibrium is performed (for details of the mapping, see\citer{nceps18} and \citers{BB90a,BB90b,BB90c} for the classical bump-on-tail paradigm). By other words, the present study, when mapped into the radial profile of fast ions in a Tokamak, shows that the temporal mesoscale radial transport is not satisfactory addressed by a QL code. This means that all phenomena which could be relevant during this phase of the evolution must be described out of the QL paradigm, at least by a correction of the instantaneous growth rate, due to the radial gradients of the thermodynamic parameter of the plasma. When using a QL code, the instantaneous growth of the spectrum is significantly underestimated in the field saturation regime and then relevant features of the fast ions transport could not be adequately reproduced.

\acknowledgments
This work has been partly carried out within the framework of the EUROfusion Consortium [Enabling Research Projects: NAT (AWP17-ENR-MFE-MPG-01), MET (CfP-AWP19-ENR-01-ENEA-05)] and has received funding from the Euratom research and training programme 2014-2018 and 2019-2020 under grant agreement No 633053. The views and opinions expressed herein do not necessarily reflect those of the European Commission.

%\bibliographystyle{_style}
%\bibliography{PLASMA}

\begin{thebibliography}{10}


\bibitem{ZCrmp}
L.~{Chen} and F.~{Zonca}, \emph{Rev. Mod. Phys.} \textbf{88}, 015008 (2016).

\bibitem{BB90a}
H.L. {Berk} and B.N. {Breizman}, \emph{Phys. Fluids B} \textbf{2}, 2226 (1990).

\bibitem{BB90b}
H.L. {Berk} and B.N. {Breizman}, \emph{Phys. Fluids B} \textbf{2}, 2235 (1990).

\bibitem{BB90c}
H.L. {Berk} and B.N. {Breizman}, \emph{Phys. Fluids B} \textbf{2}, 2246 (1990).

\bibitem{EEbook}
Y.~{Elskens} and D.F. {Escande}, \emph{Microscopic Dynamics of Plasmas Chaos}, Taylor Francis Ltd (2003).

\bibitem{AEE98}
M.~{Antoni}, Y.~{Elskens} and D.F. {Escande}, \emph{Phys. Plasmas} \textbf{5},
  841 (1998).

\bibitem{EE08}
D.F. {Escande} and Y.~{Elskens}, \emph{arXiv:0807.1839}  (2008).

\bibitem{CFMZJPP}
N.~{Carlevaro}, M.V. {Falessi}, G.~{Montani} and F.~{Zonca}, \emph{J. Plasma
  Phys.} \textbf{81}, 495810515 (2015).

\bibitem{ncentropy}
N.~{Carlevaro}, A.V. {Milovanov}, M.V. {Falessi}, G.~{Montani}, D.~{Terzani}
  and F.~{Zonca}, \emph{Entropy} \textbf{18}, 143 (2016).

\bibitem{OM68}
T.M. {O'Neil} and J.H. {Malmberg}, \emph{Phys. Fluids} \textbf{11}, 1754
  (1968).

\bibitem{OWM71}
T.M. {O'Neil}, J.H. {Winfrey} and J.H. {Malmberg}, \emph{Phys. Fluids}
  \textbf{14}, 1204 (1971).

\bibitem{MK78}
H.E. {Mynick} and A.N. {Kaufman}, \emph{Phys. Fluids} \textbf{21}, 653 (1978).

\bibitem{TMM94}
J.L. {Tennyson}, J.D. {Meiss} and P.J. {Morrison}, \emph{Physica D}
  \textbf{71}, 1 (1994).

\bibitem{KV14}
C.~{Krafft} and A.~{Volokitin}, \emph{Eur. Phys. J. D} \textbf{68}, 370 (2014).

\bibitem{L72}
M.B. {Levin}, M.G. {Lyubarski{\v i}}, I.N. {Onishchenko}, V.D. {Shapiro} and
  V.I. {Shevchenko}, \emph{Sov. Phys. JEPT} \textbf{35}, 898 (1972).

\bibitem{Pines}
W.E. {Drummond} and D.~{Pines}, \emph{Nucl. Fusion Suppl. Part.} \textbf{3}, 1049 (1962).

\bibitem{MT64}
D.C. Montgomery and D.A. Tidman, \emph{Plasma Kinetic Theory}, McGraw-Hill (1964).

\bibitem{KT73}
N.A. Krall and A.W. Trivelpiece, \emph{Principles of Plasma Physics}, McGraw-Hill (1973).

\bibitem{D82}
R.C. Davidson,  \emph{Methods in non-linear Plasma Theory}, Pergamon Press Ltd (1982).

\bibitem{pesme94}
D.~{Pesme}, \emph{Phys. Scr. T} \textbf{50}, 7 (1994).

\bibitem{Laval99}
G.~{Laval} and D.~{Pesme}, \emph{Plasma Phys. Control. Fusion} \textbf{41},
  A239 (1999).

\bibitem{GG12}
K.~{Ghantous}, N.N. {Gorelenkov}, H.L. {Berk}, W.W. {Heidbrink} and M.A. {Van
  Zeeland}, \emph{Phys. Plasmas} \textbf{19}, 092511 (2012).

\bibitem{GBG14}
K. Ghantous, H.L. Berk and N.N. Gorelenkov, \emph{Phys. Plasmas} \textbf{21}, 032119 (2014).

\bibitem{BBFW95}
H.L. Berk, B.N. Breizman, J. Fitzpatrick and H. V. Wong, \emph{Nucl. Fusion} \textbf{35}, 1661 (1995).

\bibitem{spb16}
M.~{Schneller}, Ph. {Lauber} and S.~{Briguglio}, \emph{Plasma Phys. Control.
  Fusion} \textbf{58}, 014019 (2016).
 
\bibitem{DS68}
J.M. Dawson and R. Shanny, \emph{Phys. Fluids} \textbf{11}, 1506 (1968).

\bibitem{MN69}
R.L. Morse and C.W. Nielson, \emph{Phys. Fluids} \textbf{12}, 2418 (1969).

\bibitem{D90}
C.T. Dum, \emph{J. Geophys. Res.} 95, 8095 (1990).

\bibitem{H55}
F.H. Harlow, \emph{A machine calculation method for hydrodynamic problems}, Technical report (Los Alamos Sci. Lab.) (1955).

\bibitem{HE88}
R.W. Hockney and J.W. Eastwood, \emph{Computer Simulation Using Particles}, IOP Publishing (1988).

\bibitem{D94}
C.T. Dum, \emph{Phys. Plasmas} \textbf{1}, 1821 (1994).

\bibitem{EB15}
A.A. Efimova et al., \emph{AIP Conf. Proc.} \textbf{1684}, 100001 (2015).

\bibitem{BDE17}
E.A. Berendeev, G.I. Dudnikova, and A.A. Efimova, \emph{AIP Conf. Proc.} \textbf{1895}, 120002 (2017).

\bibitem{DF14}
W. Deng and G.Y. Fu, \emph{Computer Phys. Comm.} \textbf{185}, \textbf{96} (2014).

\bibitem{LP83}
G.~{Laval} and D.~{Pesme}, \emph{Phys. Fluids} \textbf{26}, 66 (1983).

\bibitem{nceps16}
N.~{Carlevaro}, G.~{Montani}, X.~{Wang} and F.~{Zonca}, \emph{\emph{in} 43rd
  EPS Conference on Plasma Physics} \textbf{40A}, P5.018 (2016).

\bibitem{oneil65}
T.M. {O'Neil}, \emph{Phys. Fluids} \textbf{8}, 2255 (1965).

\bibitem{LP81}
E.M. {Lifshitz} and L.P. {Pitaevskii}, \emph{Course of Theoretical Physics 10: Physical Kinetics}, Butterworth-Heinemann (1976).

\bibitem{nceps18}
N.~{Carlevaro}, G.~{Montani} and F.~{Zonca}, \emph{\emph{in} 45th EPS
  Conference on Plasma Physics} \textbf{42A}, P5.1067 (2018).

\bibitem{AK66}
L.M. {Al'Tshul'} and V.I. {Karpman}, \emph{Sov. Phys. JEPT} \textbf{22}, 361
  (1966).
  
\bibitem{Klimontovich}
V.L. Klimontovich, \emph{Kinetic Theory of Nonideal Gases and Nonideal Plasmas}, pag.212, Pergamon Press Ltd (1982).

\bibitem{K69}
W.L. Kruer, J.M. Dawson, \emph{Phys. Rev. Lett.} \textbf{25}, 838 (1969).

\end{thebibliography}

\end{document}